\documentclass[a4paper,11pt]{article}
\usepackage{pos}

\def\ba{\begin{eqnarray}}
\def\ea{\end{eqnarray}}

\usepackage{color}
\newcommand{\B}{\color{blue}}

\title{Timelike electromagnetic form factors of hyperons at large $q^2$}

\author*[a]{G.~Ramalho}
\author[b,c]{M.~T.~Pe\~na}
\author[d]{K.~Tsushima}
\author[a]{Myung-Ki Cheoun}

\affiliation[a]{Department of Physics and OMEG Institute, Soongsil University, \\
  Seoul 06978, Republic of Korea}

\affiliation[b]{LIP, Laborat\'orio de Instrumenta\c{c}\~ao e F\'{i}sica 
Experimental de Part\'{i}culas, \\
Avenida Professor Gama Pinto, 1649-003 Lisboa, Portugal}
\affiliation[c]{Departamento de F\'{i}sica e Departamento de Engenharia e Ci\^encias Nucleares, 
Instituto Superior T\'ecnico,  \\  
Universidade de Lisboa, 
Avenida Rovisco Pais, 1049-001 Lisboa, Portugal}

\affiliation[d]{Laborat\'orio de 
  F\'{i}sica Te\'orica e Computacional -- LFTC, \\
  Programa de P\'osgradua\c{c}\~ao em F\'{i}sica Computacional, 
Universidade Cidade de  S\~ao Paulo,  \\
01506-000,  S\~ao Paulo, SP, Brazil}

\emailAdd{gilberto.ramalho2013@gmail.com}

\abstract{
  In the last few years there has been considerable progress in the study of the electromagnetic form factors of baryons in the timelike region, through electron-positron scattering, with increasing squared transfer momentum $q^2$. 
  The modulus of the electric ($G_E$) and magnetic ($G_M$) form factors has been measured for nucleons, hyperons and other baryons at BaBar, CLEO, Belle and BESIII.
  The novel measurements motivated the extension of a covariant quark model, developed to the spacelike region ($q^2 \le 0$), to the timelike region, without any further parameter fitting.
  The extension is based on asymptotic relations derived from analyticity and unitarity, valid for the large-$q^2$ region.
  We use the model to make predictions for the effective form factor $|G|$ (combination of $G_E$ and $G_M$) and the ratio $|G_E/G_M|$ for spin 1/2 hyperons at large $q^2$ (above 10 GeV$^2$).
  Our calculations are in good agreement with the data from CLEO and BESIII for $\Lambda$, $\Sigma^+$ and $\Xi^-$ above $q^2=15$ GeV$^2$.
  Upcoming data for $\Sigma^0$, $\Sigma^-$ and $\Xi^-$ at large $q^2$ may be used to further test our predictions.
  We also compare our model calculations with the scarce available data for $|G_E/G_M|$.
We conclude that the present $q^2$ range is not large enough to test our calculations, but that a more definitive test can be made by experiments above $q^2=20$ GeV$^2$.}

\FullConference{The 21st International Conference on Hadron Spectroscopy and Structure (HADRON2025)\\
 27 - 31 March, 2025\\
Osaka University, Japan\\}

\begin{document}
\maketitle

\section{Introduction and formalism}

The study of the internal structure of the baryons has been dominated by scattering of electrons on nucleon targets (spacelike region) in modern facilities like Jefferson Lab, MAMI and MIT-Bates~\cite{PPNP2024,NSTAR}.
In the last decades, there was a significant progress in the experimental study of the electromagnetic structure of the baryons in the timelike region (invariant square transfer momentum $q^2 > 0$), based on electron-positron collisions in facilities like BaBar, CLEO, Belle and BES~\cite{Hyperons12,Review1,Dobbs17a,BESIII19a,BESIII23a,Review2}.
The measurements of the total and differential cross sections, on the annihilation reactions $e^+ e^- \to B \bar B$, where $B$ is a spin 1/2 baryon, in a region where the invariant square transfer momentum $q^2 =s \ge 4 M_B^2$ ($M_B$ is the mass of the baryon) are used to extract information about the electromagnetic structure of baryons.
In that region, the electromagnetic form factors are complex functions of $q^2$.
Experiments measure nowadays the modulus of the electric ($G_E$) and magnetic ($G_M$) form factors, that can be combined to define the effective form factor $|G (q^2)|$, and the electromagnetic ratio $R(q^2)$~\cite{Hyperons12,Review1}
\ba
|G(q^2)|^2 = \frac{2 \tau |G_M(q^2)|^2 + |G_E(q^2)|^2}{1+ 2 \tau},
\hspace{1.5cm}
R(q^2) = \frac{|G_E(q^2)|}{|G_M(q^2)|}.
\ea
Timelike experiments reveal information about the distribution of charge and magnetism inside of hyperons  that cannot be obtained by spacelike experiments (short lifetime of hyperons).
Measurements of the effective form factors $|G|$ for several baryons have been performed at BaBar, CLEO, BESIII and Belle.
Of particular interest have been the measurements at CLEO and BESIII above $q^2=10$ GeV$^2$~\cite{Dobbs17a,Lambda1,SigmaP1,XiM1}.

More recently, it was also possible to measure the polarization of the baryon and antibaryon final states, allowing the determination of the relative phase between $G_E$ and $G_M$: $\Delta \Phi$ at BESIII for the $\Lambda$ and $\Sigma^+$~\cite{BESIII19a,BESIII23a}.
With the knowledge of $|G_M|$, $|G_E|$ and $\Delta \Phi$, (also function of $q^2$), one can write:
\ba
\frac{G_E}{G_M} = \frac{|G_E|}{|G_M|} e^{i \Delta \Phi},
\hspace{2cm}
{\rm Re}  \left( \frac{G_E}{G_M} \right)
    = 
    \frac{|G_E|}{|G_M|} \cos \Delta \Phi.
    \label{eqGEGM}
\ea

Aiming for the interpretation of the recent data on $|G|$ and $R$, we use a covariant quark model~\cite{Nucleon,Spectator,NSTAR2017}, developed for the study of electromagnetic transitions between baryon states in the spacelike region, to make predictions for the large $q^2$ timelike region.
The model takes into account both the effects of valence quarks and the excitations of the meson cloud which dresses the bare baryons.
The extension to the timelike region is made using asymptotic relations based on unitarity and analyticity, valid in the large $q^2$ region~\cite{Hyperons12}.

\section{Numerical results}

The present calculations are based on the covariant quark model formalism~\cite{Nucleon,Spectator}.
The model was developed originally for the study of transitions of the nucleon
and nucleon resonances~\cite{NSTAR2017,nstars}
and transitions between baryon states in the spacelike region
($Q^2 = - q^2 \ge 0$)~\cite{Octet4,OctetDecuplet,Baryons}.
The model has also been used in the study of inelastic timelike transitions~\cite{OctetDecuplet2,Dalitz} as well as the electroweak structure of baryons in vacuum and in a nuclear medium~\cite{Axial,Medium} and the nucleon deep inelastic scattering~\cite{Nucleon,NucleonDIS}.

In the covariant spectator quark model the electromagnetic transition current is determined using relativistic impulse approximation for the photon interaction with a single quark, while integrating over the degrees of freedom of the spectator quark-pair~\cite{Nucleon,Spectator,NSTAR2017,OctetDecuplet}.
The quark internal structure associated with the gluon and quark-antiquark dressing is parametrized by constituent quark form factors. 
The quark-diquark radial wave functions are determined phenomenologically using lattice QCD data, based on the SU(3) flavor symmetry structure that is broken explicitly.
In the low-$Q^2$ region, we consider also
effective descriptions of the meson cloud dressing of the baryon states~\cite{Spectator,OctetDecuplet}.

In the following, we use the extension of the model for the timelike region, as discussed in Refs.~\cite{Hyperons12}, to calculate $G_E$ and $G_M$.
For the extension we consider the finite correction to the asymptotic reflection relation between spacelike and timelike form factors: $G_\ell (q^2) \equiv G_\ell^{\rm SL} (q^2 - 2 M_B^2)$, where $\ell =E,M$, and $G_\ell^{\rm SL}$ are the spacelike form factors.
We use the parameters determined by the model for the octet baryons and decuplet baryons~\cite{Octet4,OctetDecuplet}.

We do not discuss the oscillatory dependence observed on the form factors of baryons because we are focused on the large $q^2$ region, where the oscillatory component is expected to be suppressed~\cite{Oscilations,BESIII-LambdaC}.

\begin{figure*}[t]
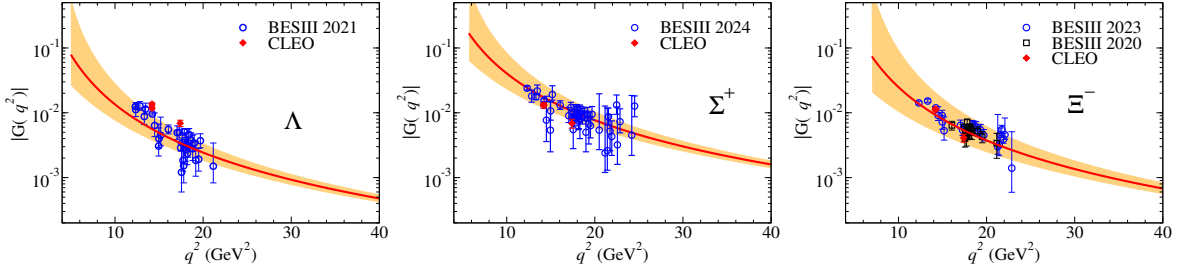

  \centerline{\mbox{
  \includegraphics[width=1.99in]{GT-Lambda-Lq2} 
  \includegraphics[width=1.99in]{GT-SigmaP-Lq2}
  \includegraphics[width=1.99in]{GT-XiM-Lq2} }}
\caption{\footnotesize{Effective form factors $|G(q^2)|$ in the large-$q^2$ region ($q^2 > 10$ GeV$^2$)~\cite{Hyperons12}.
    The data are from BESIII~\cite{Lambda1} ($\Lambda$), \cite{SigmaP1} ($\Sigma^+$),
    \cite{XiM1} ($\Xi^-$) and CLEO~\cite{Dobbs17a}.
    The colored band indicates the theoretical deviations from the asymptotic relations. 
\label{figG1}}}
\end{figure*}

In Fig.~\ref{figG1}, we compare our model calculations for $|G|$ with the $q^2> 10$ GeV$^2$ data for $\Lambda$, $\Sigma^+$, and $\Xi^-$.
In addition to the asymptotic estimate (central line), we present also the theoretical limits, based on finite corrections to the asymptotic relations (see Refs.~\cite{Hyperons12} for more details).
From the comparison, we can conclude that the model provides a good description of the data above $q^2=15$ GeV$^2$, within the theoretical limits.
It is worth mentioning that the calculations for $|G|$ were performed before the data from BESSIII, included in the figures, became available.
The predictions for $\Sigma^0$, $\Sigma^-$ and $\Xi^0$ are waiting for $q^2 > 10$ GeV$^2$ data.

The present calculations for $|G|$ suggest that the region $q^2 > 15$ GeV$^2$ is already in the range where the asymptotic behavior of the form factors can be observed.
One notices, however, that the present data may still be in the nonperturbative QCD region, and that the onset for the perturbative QCD falloff of the form factors happens in a much higher region of $q^2$~\cite{Hyperons12}.

We also use the model to calculate the ratio $|G_E/G_M|$.
At the moment, measurements of the ratio are restricted to $\Lambda$, $\Sigma^+$, and $\Lambda_c^+$~\cite{Lambda1,SigmaP1,BESIII-LambdaC}, and to moderate values of $q^2$ ($q^2 \sim 10$ GeV$^2$) for the hyperons.
Our calculations underestimate the available data for $\Lambda$ and  $\Sigma^+$ (see discussion of Refs.~\cite{Hyperons12}).
Since our calculations are based on real functions (because we use results from the spacelike region) they cannot be directly compared with the timelike data, for moderate values of $q^2$ (complex functions), but only on the large region of $q^2$ where the imaginary components of the form factors are expected to be significantly suppressed.

\begin{figure*}[h]
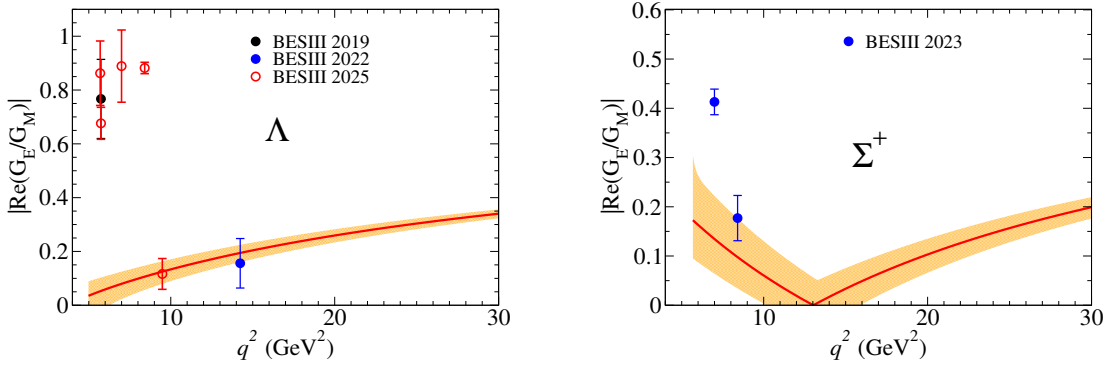

  \vspace{.3cm}
  \centerline{\mbox{
\includegraphics[width=2.6in]{Lambda-ReGEGM} \hspace{1.cm}
\includegraphics[width=2.6in]{SigmaP-ReGEGM} }}
    \caption{\footnotesize{Real part of
        the ratio $G_E/G_M$ for the $\Lambda$
        and $\Sigma^+$ hyperons. Data from
        Refs.~\cite{BESIII19a,BESIII22-25} ($\Lambda$)
        and \cite{BESIII23a} ($\Sigma^+$).
        The colored band indicates the theoretical deviations from the asymptotic relations. 
        The uncertainties associated with the phase $\Delta \Phi$ are not included in the data points.
        \label{figRatio1}}}
\end{figure*}

One notices, however, that the calculations can be compared with the real part of $G_E/G_M$, if the relative phases are known [see Eq.~(\ref{eqGEGM})].
In Fig.~\ref{figRatio1} we compare our model for the modulus of $G_E/G_M$ with the experimental value for $|{\rm Re}(G_E/G_M)|= R(q^2) |\cos \Delta \Phi|$.
  The relative angle $\Delta \Phi$ is in some cases very imprecise and very sensitive to the value of $q^2$.
For a simpler comparison with the data we omit the errors on $\Delta \Phi$ in the figures.

Figure~\ref{figRatio1} shows that our model calculations are consistent with the data above $q^2=8$ GeV$^2$. 
More definitive conclusions can only be drawn when data for larger values of $q^2$ become available, and/or when $|\cos \Delta \Phi| \approx 1$.
Nevertheless, the results from Fig.~\ref{figRatio1} hint that the onset of validity of our model for $|{\rm Re}(G_E/G_M)|$ is located in the region not so far the one planned for future experiments, which is $q^2=20$--30 GeV$^2$.
The improvement in the accuracy of the relative angle may also help to test the present predictions.

In this presentation, we reduced the discussion to elastic form factors of spin 1/2 baryons.
It is worth noticing, however, that with minor modifications, we can also study spin 3/2 systems, like the $\Omega^-$ baryon~\cite{Omega}, and the $e^+ e^- \to \Lambda \bar \Sigma^0$ reaction~\cite{BESIII-2024-Nature,Octet4}.

\section{Summary and conclusions}

We use asymptotic relations to extend the covariant spectator quark model from the spacelike region to the elastic timelike region and make predictions for the effective form factors and electromagnetic ratio of hyperons from the baryon octet.
We also estimate the theoretical uncertainties of the calculations in the region of study.

In general, we obtain a good agreement with the effective form factor $|G|$ data above $q^2=15$ GeV$^2$.
As for the function $R= |G_E/G_M|$,  our calculations are expected to be accurate only for larger values of $q^2$, when the real part of the form factors dominates.
The comparison of our calculations for the real part of $G_E/G_M$ with the data for $\Lambda$ and $\Sigma^+$ suggests that our predictions may be tested in experiments above $q^2= 20$ GeV$^2$.
New accurate data from BESIII, Belle or PANDA for $|G|$, $|G_E/G_M|$ and  $\Delta \Phi$ may be used in the near future to test our predictions.

The formalism developed for the SU(3) quark flavor sector can be extended in the near future to charmed baryons, including $\Lambda_c^+$.

\acknowledgments
G.R.~and M.-K.C.~were supported by the National
Research Foundation of Korea (Grant  No.~RS-2021-NR060129).
M.T.P~was supported by the Portuguese Science Foundation FCT
under project CERN/FIS-PAR/0023/2021, and FCT computing project 2021.09667.
K.T.~was supported by CNPq, Brazil, Processes 
No.~304199/2022-2, FAPESP Process No.~2023/07313-6,
and by Instituto Nacional de Ci\^{e}ncia e Tecnologia -- Nuclear Physics and Applications
(INCT-FNA), Brazil, Process No.~464898/2014-5.

\end{document}